\documentclass{emulateapj}

\usepackage{amsmath}
\usepackage{epsfig}
\usepackage{amssymb}
\usepackage{graphicx}
\usepackage{color}
\usepackage{natbib}

\usepackage{apjfonts} 
\usepackage{amsmath,amstext}
\usepackage[breaklinks,colorlinks,citecolor=blue,linkcolor=magenta]{hyperref} 
\usepackage[all]{hypcap} 
\usepackage{verbatim}

\begin{document}

\title[Extreme Pop III Star Formation]{Extreme Primordial Star Formation Enabled by High Redshift Quasars}

\author{Jarrett L. Johnson\altaffilmark{1} and Aycin Aykutalp\altaffilmark{2}}

\altaffiltext{1}{XTD-NTA, Los Alamos National Laboratory, Los Alamos, NM 87545}
\altaffiltext{2}{XTD-IDA, Los Alamos National Laboratory, Los Alamos, NM 87545}

\begin{abstract}
High redshift quasars emit copious X-ray photons which heat the intergalactic
medium to temperatures up to $\sim$ 10$^6$ K.  At such high
temperatures the primordial gas will not form stars until it is
assembled into dark matter haloes with masses of up to $\sim$ 10$^{11}$
M$_{\odot}$, at which point the hot gas collapses and cools under
the influence of gravity.  Once this occurs, there is a 
massive reservoir of primordial gas from which stars can form,
potentially setting
the stage for the brightest Population (Pop) III starbursts in the early
Universe.  Supporting this scenario, recent observations of quasars
at z $\sim$ 6 have revealed a lack of accompanying
Lyman $\alpha$ emitting galaxies, consistent with suppression of
primordial star formation in haloes with masses below $\sim$ 10$^{10}$
M$_{\odot}$.  Here we model the chemical and thermal evolution of the
primordial gas as it collapses into such a massive halo irradiated by
a nearby quasar in the run-up to a massive Pop III starburst.
We find that within $\sim$ 100 kpc of the highest redshift
quasars discovered to date the Lyman-Werner
flux produced in the quasar host galaxy may be high
enough to stimulate the formation of a direct collapse black hole
(DCBH).  A survey with single
pointings of the NIRCam instrument at individual known high-z quasars
may be a promising
strategy for finding Pop III stars and DCBHs with the {\it James Webb
  Space Telescope}.
\end{abstract}


\maketitle

\section{Introduction}
The hunt for the first generation of stars, so-called Population (Pop)
III stars, is generally carried out following two different approaches.  In the first,
long-lived extremely metal-poor stars are sought in large surveys of
our Galaxy, in order to place constraints on the initial mass
function of primordial stars (e.g. Frebel \& Norris 2015; Hartwig et
al. 2019).  In the second, deep
observations are made of the distant, high redshift Universe, in order
to identify galaxies that may host metal-free star
formation (e.g. Inoue 2011; Sobral et al. 2015; Xu et al. 2016).  The first
approach will only succeed in uncovering Pop III stars if they are
sufficiently low-mass that they are still burning their nuclear fuel
in the Milky Way today (or in nearby dwarf galaxies; Magg
et al. 2018).  The second approach will only succeed if Pop
III star-hosting galaxies are bright enough to be detected in deep
surveys of the early Universe (e.g. Jaacks et al. 2018; Sarmento et al. 2018).  With the sensitivity of these
surveys on the verge of great improvements, for example with the
launch of the {\it James Webb Space Telescope} (JWST; e.g. Gardner et
al. 2006; Zackrisson et al. 2012; Kalirai 2018), it is key to make predictions for how Pop
III galaxies may be discovered.

  \begin{figure*}
   \includegraphics[angle=0,width=7.0in]{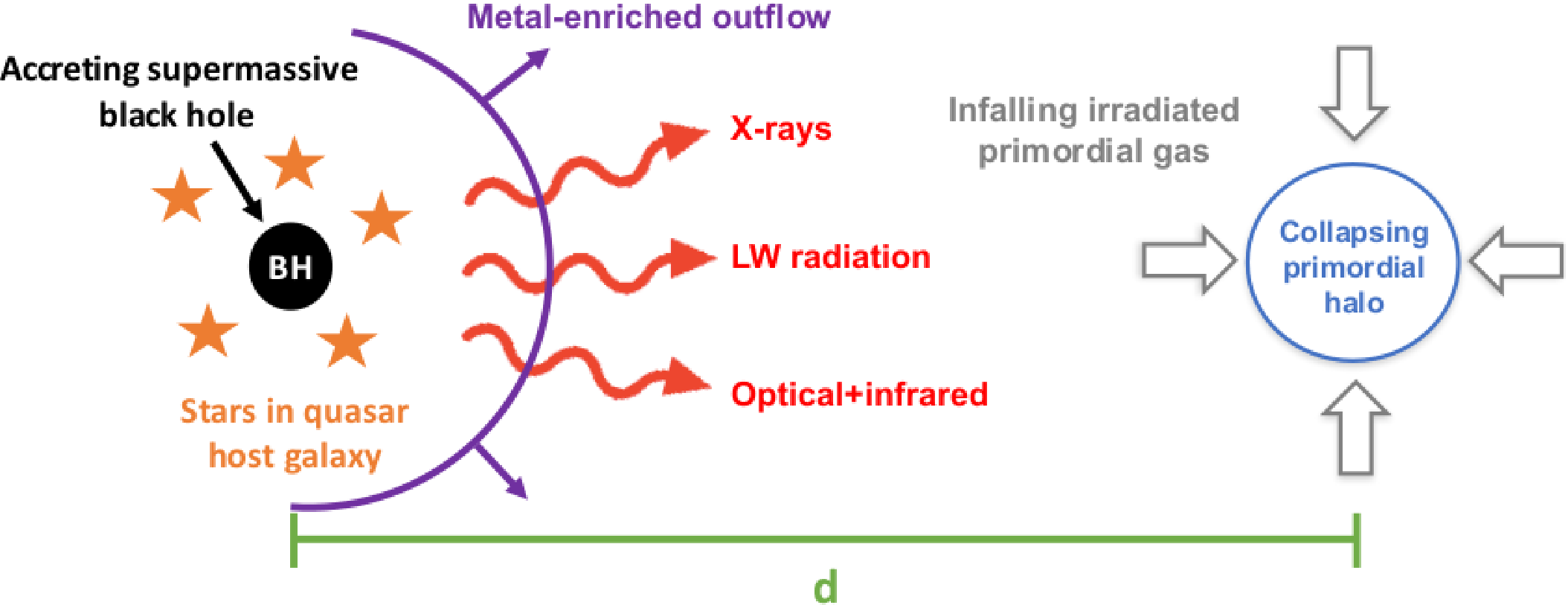}
   \caption{Schematic of the scenario we consider, the evolution of
     collapsing primordial gas ({\it right}) under the influence of X-rays, LW
     and optical/infrared radiation from a quasar ({\it left}) composed of an acceting black hole
     and stars in its host galaxy at
     redshift $z$ $\ga$ 6.  The intense X-ray heating of the
     primordial gas
     prevents its collapse into its host halo until the halo mass exceeds
     $\sim$ 10$^{10}$ M$_{\odot}$ (see Figure 2), resulting in the formation of a massive cluster of
     Population III stars or a direct collapse black hole.}
   \end{figure*}

The brightest Pop III galaxies are likely to form in the largest dark
matter (DM) haloes which harbor metal-free gas in the early Universe.
Such galaxies are expected to form in regions where Pop III star formation
is delayed until a large halo is assembled, either due to radiative
feedback (e.g. O'Shea \& Norman 2008; Trenti et al. 2009; Johnson
2010; Safranek-Shrader et al. 2012; Xu et al. 2013) or a violent merger history (Inayoshi et al. 2018).
In the case of the former, high host halo masses are achieved
when the primordial gas is subject to a stellar ionizing radiation field
which maintains its temperature at $\ga$ 10$^4$ K and prevents its collapse until a
halo with such a high virial temperature, corresponding to a mass of
up to $\sim$ 10$^9$ M$_{\odot}$, is assembled (e.g. Dijkstra et
al. 2004; Noh \& McQuinn 2014; Visbal
et al. 2017; Yajima \& Khochfar 2017).

While haloes of a billion solar masses are much larger than those
thought to host the very first stars at $z$ $\sim$ 20 (e.g. Greif
2015), it is possible that even more massive haloes may also host
primordial star formation in regions where the gas is heated to even
higher temperatures.  In particular, the highest temperatures of the
intergalactic medium (IGM) are likely to arise in the vicinity
of bright quasars powered by accretion of gas onto supermassive black holes
at $z$ $\ga$ 6 (e.g. Mortlock et al. 2011; Wu et al. 2015; Ba{\~n}ados
et al. 2017; Pons et al. 2018; Yang et al. 2018).  The intense photoheating by the copious X-rays
emitted from these objects can drive the primordial gas to
temperatures of up to $\sim$ 10$^6$ K (e.g. Bolton et al. 2009;
Aykutalp et al. 2014; Smidt
et al. 2018).  In such environments, the gas will only collapse into
extremely large DM haloes with masses of up to $\sim$ 10$^{11}$
M$_{\odot}$ (e.g. Barkana \& Loeb 2001),
setting the stage for the brightest Pop~III starbursts in the early Universe.

There is recent obserational evidence that star formation is, in fact,
suppressed in DM haloes with masses below this range in the vicinity
of high redshift quasars (Goto et al. 2017).  In their work, Goto et al. (2017) report a dearth of Lyman
$\alpha$ emitting galaxies within several megaparsecs of a luminous quasar at
$z$ = 6.4, consistent with suppression of star formation in haloes
with masses below $\sim$ 10$^{10}$ M$_{\odot}$ due to the strong X-ray
heating of the IGM by the quasar (see also Simpson et al. 2014; Ota et
al. 2018; Uchiyama et al. 2018).\footnote{Alternatively, if there is an underdensity of DM haloes
in the vicinity of high redshift quasars this could also explain the
observed trend (but see e.g. Balmaverde et al. 2017; Ginolfi et al. 2018).}  This finding supports the tantalizing
possibility that the strong radiative feedback from high redshift
quasars may delay primordial star formation until extremely massive
haloes are assembled, suggesting that the conditions for the brightest
Pop~III starbursts may indeed occur in the neighborhoods of bright high
redshift quasars.

Here we study the evolution of primordial gas irradiated by a luminous
quasar at high redshift, in order to
predict the nature of the primordial objects formed during its
collapse.  In Section 2 we lay out the methodology of our
calculations.  In Section 3 we present our basic results
illustrating the evolution of irradiated gas, while in Section 4 we
explore the nature of the objects that are formed in its collapse.  In
Section 5 we demonstrate the validity of this model, as it depends on
the suppression of star formation in haloes neighboring high redshift
quasars during their full growth history.  In Section 6 we present
predicted fluxes for these objects and outline a promising strategy
for using the JWST to search for them.  Finally, we conclude in
Section 7 with a brief summary of our findings.

\section{Methodology}
The basic scenario that we consider is illustrated schematically in
Figure 1.  On the left, a quasar at high redshift (e.g. $z$ $\ga$ 6) emits
radiation in three distinct wavebands that impact the evolution of the
primordial gas collapsing into the massive DM halo, on the right, that
sits at a
distance $d$ from the quasar.  A portion of this radiation is emitted
during the accretion process of the central black hole and a portion
is from the stellar population in the quasar host halo.  In addition,
some fraction of the metals produced by the stars inhabiting the host
halo are assumed to be entrained in a galactic wind driving an outflow.

The accreting black hole powering the
quasar emits high energy X-ray radiation, most of which 
escapes its host galaxy, although a small portion is converted into
Lyman-Werner (LW) H$_{\rm 2}$-dissociating radiation that escapes the
host galaxy.
Here we assume that the X-rays which escape the host galaxy are
monoenergetic at 1 keV, which is in the range in which most X-ray
energy has been observed to escape from high-z quasars (e.g. Nanni et
al. 2017).\footnote{We also note that Smidt et al. (2018) find good
  agreement with the available data on the Mortlock et al. (2011)
  quasar at $z$ = 7.1 by adopting such a monoenergetic X-ray spectrum
  in a fully cosmological radiation hydrodynamics calculation.}  For the flux of LW radiation $J_{\rm 21, BH}$ at the location of the
target primordial halo that is produced due to reprocessing of the
radiative energy emitted in the accretion process, as well as due to
diffuse emission in the host galaxy, we assume a simple scaling that
has been derived from post-processing of cosmological radiation hydrodynamics calculations
(Barrow et al. 2018):

\begin{equation}
 J_{\rm 21, BH} = 80 \, \left(\frac{M_{\rm BH}}{10^8 \, {\rm
       M_{\odot}}}   \right)  \left(\frac{f_{\rm Edd}}{1}   \right)
 \left(\frac{d}{1 \, {\rm Mpc}}    \right)^{-2} \mbox{\ ,}
\end{equation}
where $d$ is the distance between the quasar and the target primordial
halo (as shown in Figure 1), $M_{\rm BH}$ is the mass of the black
hole powering the quasar, and $f_{\rm Edd}$ is the ratio of the X-ray
luminosity to the
Eddington luminosity.  Finally, the units of $J_{\rm 21, BH}$ are the standard 10$^{-21}$
erg s$^{-1}$ cm$^{-2}$ Hz$^{-1}$ sr$^{-1}$.  The 
X-ray energy that escapes the quasar host halo propagates into the IGM,
where it heats the primordial gas to high temperatures.  This is shown
in Figure 2, which illustrates schematically the impact of radiation
emitted from a high redshift quasar on the primordial gas in the
surrounding IGM.   

The stellar population in the quasar host galaxy is assumed to
contribute to the LW flux impinging on the target primordial halo, as
well as to a flux of optical and infrared radiation at energies $\ge$
0.75 eV that can destroy H$^-$, an important precursor to the
formation of H$_{\rm 2}$ in the primordial gas.  The LW flux $J_{\rm
  21, *}$  produced
by the stellar population is expressed in terms of the stellar mass
$M_{\rm *}$ in
the quasar host halo, as follows (Johnson et al. 2013):

\begin{equation}
 J_{\rm 21, *} = 60 \, \left(\frac{M_{\rm *}}{10^{10} \, {\rm
       M_{\odot}}}   \right)  
 \left(\frac{d}{1 \, {\rm Mpc}}    \right)^{-2} \mbox{\ .}
\end{equation}
This rate has been chosen to correspond to an effective stellar
temperature of $T_{\rm *}$ = 3 $\times$ 10$^4$ K, in line with the cosmological
average stellar properties at $z$ $\sim$ 6 presented in Agarwal \&
Khochfar (2015).  Also following these authors, we have adopted this effective stellar
temperature in evaluating the rate of H$^-$ destruction by optical and
infrared radiation, although in Section 3 we explore the impact on our
results of assuming either older (cooler) or younger (hotter) stellar
populations (e.g. Shang et al. 2010).  

To model the evolution of the primordial gas as it collapses into the
massive primordial target halo (shown at right in Figure 1), we use these radiative fluxes
in the one-zone primordial chemistry code adopted in Johnson \&
Dijkstra (2017).  To properly incorporate the effects of X-rays in
this model, we have made the following five modifications.  
(1) We calculate the rate of photoionization of both hydrogen and
helium species, using the cross sections presented in Osterbrock \&
Ferland (2006).  
(2) We account for the partitioning of photoelectron energy into secondary
ionizations and collisional heating, as described by Shull \& van Steenberg
(1985).  
(3) We adopt an approximate treatment for the local attenuation of the
X-ray flux in the target primordial halo, following Johnson et al
(2014).
(4) We account for Compton heating of the primordial gas due to
X-rays.
(5) We initialize our calculations at a hydrogen number density of
0.01 cm$^{-3}$, corresponding to roughly a factor of two lower than the
virial density at $z$ $\simeq$ 6.\footnote{We have tested the
  sensitivity of our results to the choice of initial density.  While for
  lower initial densities the free electron fraction is found to be
  slightly higher due to the lower recombination rate, we find
  essentially identical results at the high densities at which the
  final fate of the collapsing gas is determined.}

It is important to note that this treatment is appropriate for
primordial gas that starts at the low densities characterizing the
high-z IGM, as has been extensively studied (e.g. Dijkstra et
al. 2004; Okamoto et al. 2008; Johnson et al. 2014;  Chon \& Latif
2017; Wu et al. 2019).  This is distinct
from the situation in which gas is pre-collapsed in haloes and is
photoevaporated by ionizing photons, as studied in detail by
e.g. Iliev et al. (2005).  We emphasize, however, that the impact of
X-rays in potentially stimulating primordial gas cooling through the
creation of free electrons that catalyze H$_{\rm 2}$ formation is
included in our one-zone model (as also in e.g. Inayoshi
\& Omukai 2011).

As we show in the next Section, our results are in basic
agreement with those gleaned from the similar
framework presented in Inayoshi \& Omukai (2011) and in the
cosmological calculations presented in Regan et al. (2016), although in many
cases the X-ray fluxes that we consider are much higher than those considered
by these authors.

  \begin{figure}
   \includegraphics[angle=0,width=3.2in]{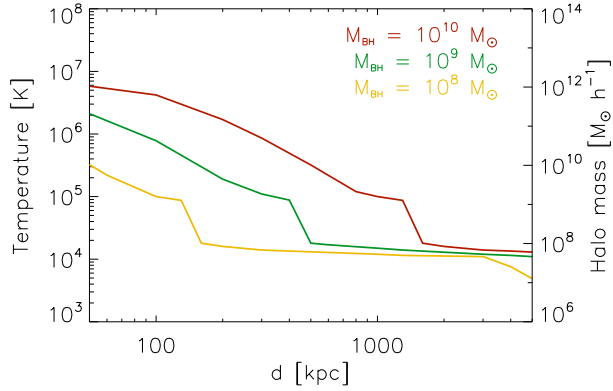}
   \caption{Temperature of the primordial gas ({\it left axis}), as a function of
     distance $d$ from quasars powered by Eddington accretion onto black
     holes of three distinct masses:  10$^8$, 10$^9$ and 10$^{10}$
     M$_{\odot}$.  The gas is assumed to be at the
     cosmic mean density at $z = 6$.  The kinks in the curves are due
     to recombination of helium and hydrogen ions and the additional
     cooling that they provide.  Also shown is the minimum halo mass required for
     the primordial gas to collapse ({\it right axis}) to form
     either stars or a DCBH.}
   \end{figure}

\begin{figure*}
   \includegraphics[angle=0,width=7.0in]{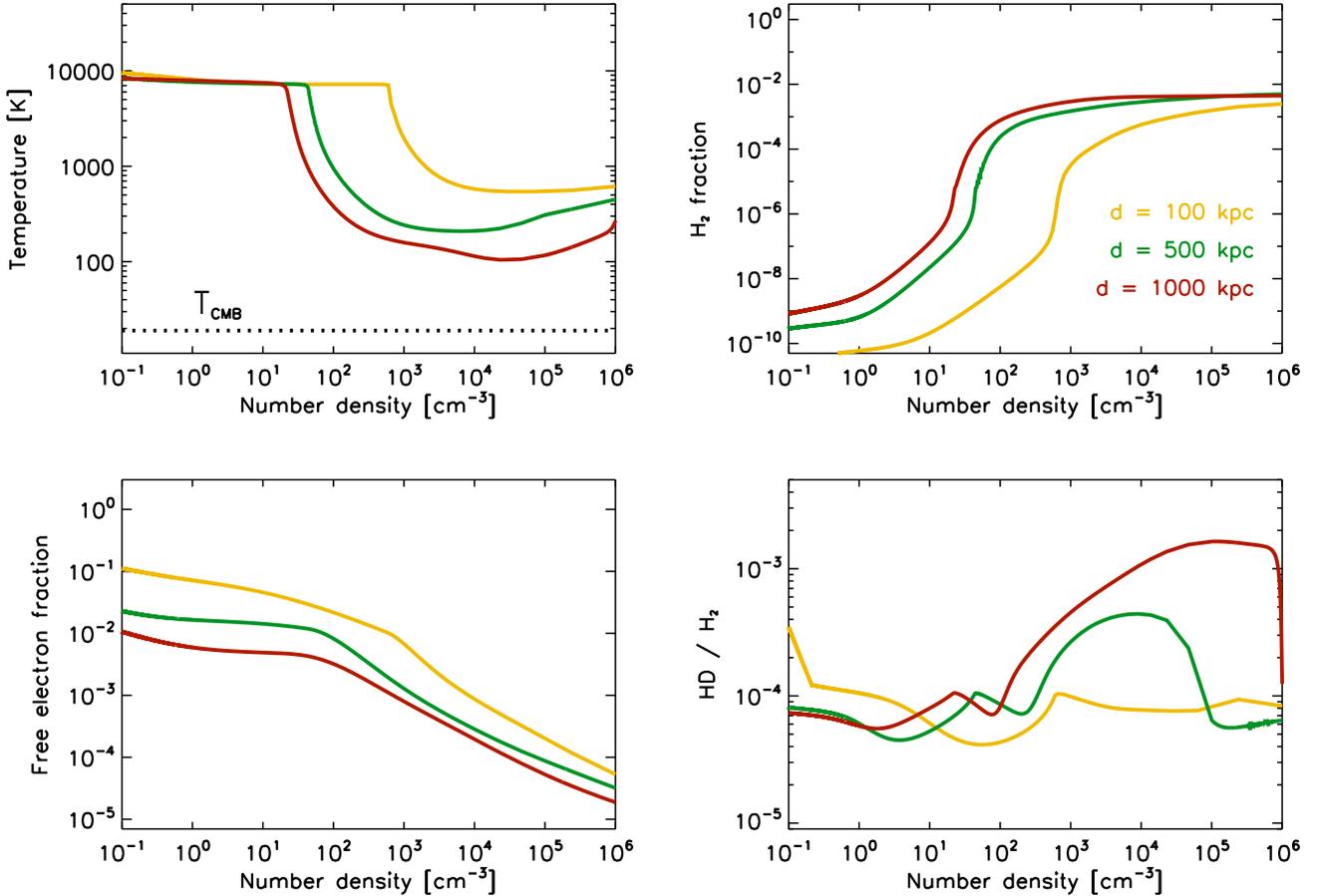}
   \caption{Evolution of the collapsing primordial gas at
   three distinct distances $d$ from a quasar powered by Eddington
   accretion onto a 10$^{8}$ M$_{\odot}$ black hole within a
   host galaxy containing 10$^{10}$ M$_{\odot}$ in stars: 100 
   ({\it yellow}), 500 ({\it green}), 1000 kpc ({\it red}).   Clockwise from the
   top-right panel are shown, as functions of the number density to which the
   gas has collapsed, the following:  the H$_{\rm 2}$ fraction, the
   ratio of the HD and
   H$_{\rm 2}$ number densities in the gas, the free electron fraction and the gas
   temperature.  The temperature floor set by the cosmic microwave
   background (CMB) at $z$
 = 6 is shown by the horizontal dotted line in the top-left
 panel. }
   \end{figure*}

\section{Evolution of Strongly Irradiated Primordial Gas}
Given the extreme radiative environment in the vicinity of a luminous
quasar in the early Universe, the evolution of the primordial gas
exposed to this radiation is strongly dependent on its proximity to
the quasar.  In addition, the chemistry of the primordial gas is
impacted in a complex manner by the various types of radiation that
are emitted from the accreting black hole and its surrounding host
galaxy.  In particular, while the LW radiation and the
optical/infrared radiation generally act to suppress the abundance of
the key molecular coolant H$_{\rm 2}$, the intense X-ray radiation
acts to produce free electrons which stimulate its formation
(e.g. Glover 2003; Aykutalp et al. 2014).

Figure 3 shows the evolution of the primordial gas as it collapses to
high density in a massive primordial DM halo exposed to the radiation
produced by a quasar powered by Eddington accretion onto a 10$^8$ M$_{\odot}$
supermassive black hole within a host galaxy containing 10$^{10}$
M$_{\odot}$ in stars.  While the gas is initially at very high
temperatures when it is at the density of the IGM prior to its
collapse, as shown in Figure 2, once it is bound in a sufficiently
massive DM halo it is able to collapse to high density and its
temperature then drops to the $\sim$ 10$^4$ K floor set by atomic
hydrogen cooling.  Then, depending on its distance from the quasar,
the temperature drops by up to another two orders of magnitude due to 
molecular cooling by H$_{\rm 2}$.  There are two competing
  processes which dictate the degree to which H$_{\rm 2}$ cooling
  affects the evolution of the primordial gas.  While the free electron fraction
is elevated due to photoionization of hydrogen and helium
species by the X-rays, leading to the catalyzed formation of H$_{\rm 2}$
molecules, the LW and optical/infrared radiation emitted from the quasar host galaxy also
strongly suppress the H$_{\rm 2}$ fraction.  The result is that, closer to the quasar
source, the gas remains hotter at the highest densities than it is
farther away.

The net impact of the X-ray flux on the thermal evolution of the gas
is shown in Figure 4, where the solid lines show the temperature of
the gas with the X-ray flux included in the calculation and the dashed
lines show it with them excluded.  It is clear that while they
strongly heat the gas at low densities, the X-rays 
also have the impact of enhancing molecular cooling at high
densities (e.g. Aykutalp et al. 2014; Inayoshi \& Tanaka 2015; Latif et al. 2015; Glover 2016).  Their effect
is strongest closest to the quasar source, where in their absence the
gas temperature remains elevated at $\simeq$ 10$^4$ K at a distance of
100 kpc, signifying a very strong suppression of H$_{\rm 2}$
formation.

At larger distances from the quasar source, where the gas is able to
cool due to the radiation from H$_{\rm 2}$ molecules, it is also
possible for the coolant HD to play a role in the thermal evolution of
the gas, as shown in Figure 5.  As the gas
temperature drops to lower values (at high density) with increasing distance from the
quasar source, HD cooling becomes stronger, as reflected in the higher HD 
fractions at larger distances in the bottom-right panel of Figure 3.
This is consistent with previous work showing that HD cooling can be
triggered in primordial gas with an elevated free electron fraction
(e.g. Nagakura \& Omukai 2005; Johnson \& Bromm 2006; Nakauchi et al. 2014).
Due to the lower temperatures to
which the gas is able to cool, in part due to HD cooling, farther from the quasar source, the fragmentation scale in the
primordial gas is likely smaller and the characteristic initial mass function of stars that
form may be shifted to lower masses (e.g. Uehara \& Inutsuka 2000; Ripamonti 2007; McGreer \& Bryan
2008).  

While we have assumed a cosmological average stellar population for
the quasar host galaxy in our results presented thus far, in Figure 6
we show the thermal evolution of the gas irradiated by both an older
and a younger stellar population, corresponding to characteristic
stellar radiation temperatures of $T_{\rm *}$ = 10$^4$ and 10$^5$ K, respectively.
Principally due to the elevated rate
of photodetachment of H$^-$ by infrared and optical photons, an older
stellar population clearly delays the cooling of the gas and in some
cases prevents it from cooling to temperatures much below 10$^4$ K,
consistent with previous results (e.g. Wolcott-Green et al. 2012, 2017; Sugimura et al. 2014;
Agarwal \& Khochfar 2015; Latif et al. 2015).  In contrast, the gas
evolves similarly to our fiducial case with a younger, hotter stellar
population in the quasar host halo.

\section{Direct Collapse Black Hole Formation}
The final fate of the quasar-irradiated primordial gas in 
a massive DM halo is strongly dependent on the temperature to which it
cools as it collapses, as this dictates the scale on which
gravitational fragmentation occurs.  In general, if the gas remains
hotter during its collapse then it is likely to fragment into stars
with a mass function shifted to higher masses.  In extreme cases in
which molecular cooling is strongly suppressed, the gas may remain at 
temperatures of $\simeq$ 10$^4$ K, leading to the formation
of supermassive stellar objects (e.g. Begelman 2010; Hosokawa et al. 2013;
Schleicher et al. 2013; Haemmerl{\'e} et al. 2018) with masses of 10$^4$ - 10$^6$ M$_{\odot}$
which promptly collapse into so-called direct collapse black holes
(DCBHs; for reviews see Volonteri 2012; Latif \& Ferrara 2016; Valiante et
al. 2017; Smith et al. 2017; Woods et al. 2018).  Here we explore the conditions required
for the formation of DCBHs in primordial haloes irradiated by quasars.

As shown in the previous Section, it is possible for the 
primordial gas to remain at the high temperatures required for DCBH
formation if the radiative flux from the quasar is
sufficiently strong.  Table 1 shows the maximum distance $d_{\rm
  DCBH}$ from a quasar powered by Eddington accretion out to which
DCBH formation can occur, for various masses $M_{\rm BH}$ of the black
hole powering the quasar and across a range of temperatures
characterizing the stellar population in the quasar host galaxy.  For
our fiducial case with $T_{\rm *}$ = 3 $\times$ 10$^4$ K, we find that
DCBH formation may occur within 50 kpc of a quasar powered by
Eddington accretion onto a 10$^8$ M$_{\odot}$ black hole, while it may
occur out to 500 kpc in the extreme case of one powered by a 10$^{10}$
M$_{\odot}$ black hole.  In general, the values we find for $d_{\rm
  DCBH}$ increase with decreasing stellar radiation temperature,
implying that DCBHs can form farther out from quasars hosting older stellar populations.     

  \begin{figure}
   \includegraphics[angle=0,width=3.5in]{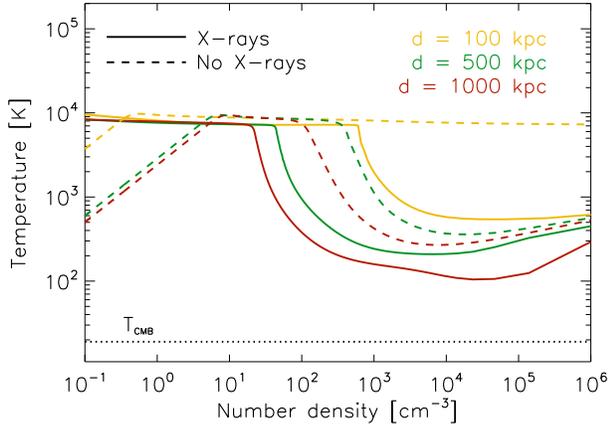}
  \caption{Like the top-left panel in Figure 3, but now showing the evolution of the
    collapsing gas with ({\it solid}) and without ({\it dashed}) the quasar source X-ray flux.}
   \end{figure}

  \begin{figure}
   \includegraphics[angle=0,width=3.5in]{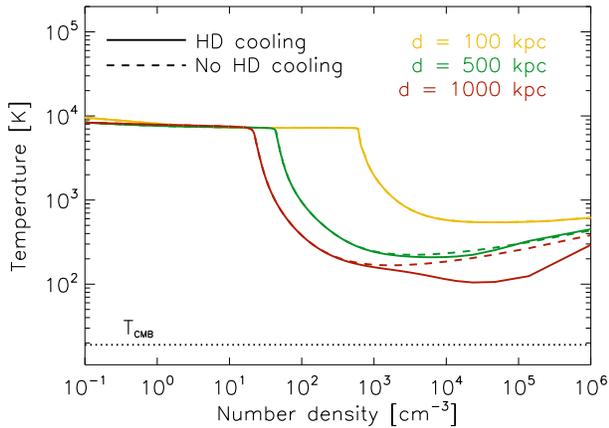}
   \caption{Like the top-left panel in Figure 3, but now showing the evolution of the
    collapsing gas with ({\it solid}) and without ({\it dashed}) HD
    cooling included.  Note that the lines corresponding to $d$
    = 100 kpc are essentially identical and overlap one another.}
   \end{figure}

  \begin{figure}
   \includegraphics[angle=0,width=3.5in]{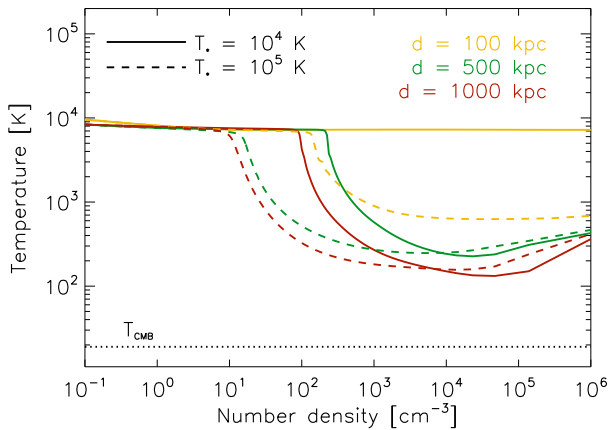}
  \caption{Like the top-left panel in Figure 3, but now showing the evolution of the
    collapsing gas under the influence of LW and optical/infrared
    radiation from a stellar population with a characteristic
    temperature of 10$^4$ K ({\it solid}) and 10$^5$ K ({\it dashed}).}
   \end{figure}

While Table 1 shows results for a fiducial choice of $M_{\rm
  *}$/$M_{\rm BH}$ = 100 for the stellar to black hole mass in the
source quasar, it is important to note that this ratio can take on
smaller values particularly in the early Universe when there has been
little time for star formation around the rapidly accreting black
holes powering quasars.  In particular, Venemans et al. (2017a,b) find
evidence for values as low as $M_{\rm *}$/$M_{\rm BH}$ $\sim$ 20 for the 
highest redshift quasars.  Furthermore, in some scenarios for the
earliest stages of growth of high-z quasars, the central black hole
mass can be even larger relative to that of the stellar population
(e.g. Agarwal et al. 2013).  Table 2 shows the values of $d_{\rm
  DCBH}$ that we find for various ratios of the stellar to black hole
mass in the quasar host halo.  In general, given the contribution that
the stellar component makes to the production of H$_{\rm
  2}$-dissociating LW radiation, we find that DCBH formation can occur
out to larger distances for a larger stellar population, at a given
value of the central black hole mass.  The results shown in Table 2
are captured well by a single fitting formula which expresses $d_{\rm
  DCBH}$ in terms of the black hole mass $M_{\rm BH}$ and stellar mass 
$M_{\rm *}$ in the quasar host halo:

\begin{equation}
d_{\rm DCBH} \simeq 30 \, {\rm kpc} \, \left( \frac{M_{\rm BH}}{10^9
    \, {\rm M_{\odot}}}    \right)^{0.5}   \left( \frac{M_{\rm *}}{M_{\rm BH}}  \right)^{0.4}    \mbox{\ ,}
\end{equation}
where this is valid under the assumption of Eddington accretion onto
the black hole and given our fiducial case of a $z$ $\sim$ 6 cosmological average stellar population
with a characteristic temperature of $T_{\rm *}$ = 3 $\times$ 10$^4$
K.  Outside of $d_{\rm DCBH}$ we expect that the gas will cool to
sufficiently low temperatures that the gas will readily fragment and
form a cluster of Pop III stars, instead of a DCBH.

While the radiation emitted from high-z quasars may provide the
conditions for the formation of these objects, such quasars are also known to emit
metal-enriched ouflows which pollute the IGM, as shown schematically
in Figures 1 and 9.  These metals, if mixed with the collapsing primordial
gas, will act to preclude the formation
of Pop~III stars and DCBHs.  At an average velocity of 100 km
s$^{-1}$ (e.g. Girichidis et al. 2016), the outflow would progress at 
most $\simeq$ 100 kpc within the age of the Universe at $z$ = 6.  This
is, in fact, broadly consistent with the results of a cosmological radiation
hydrodynamics simulation (Smidt et al. 2018) of the formation of a $z$ = 7.1 quasar matching the
observable properties of the Mortlock et al. (2011) quasar, in which the
metal-enriched region extends out to $\simeq$ 50 kpc from the compact
quasar host galaxy.\footnote{We note that a similar extent of $\simeq$ 50 kpc for
  metal-enriched outflows from quasar host galaxies at $z$ $\simeq$ 7.5 is also
  found in other recent cosmological simulations (e.g. Ni et
  al. 2018).}  In addition, this is also in line with the $\simeq$ 30
kpc extent of the outflow inferred for a bright $z$ = 6.4 quasar by
Cicone et al. (2015) as well as with the $\sim$ 10 kpc extent of
metal-enriched regions around star-forming galaxies at similar
redshifts detected with ALMA (e.g. Fujimoto et al. 2019).  As these distances are comparable to those we find
for $d_{\rm DCBH}$ around quasars powered by relatively small black
holes and with low stellar to black hole mass ratios, it appears
that the rate of DCBH formation around high-z quasars is likely to be
somewhat reduced due to metal enrichment of the primordial IGM in
their vicinity.  Given this, the Pop~III galaxies which
we expect to form farther out (i.e. at $d$ $>$ $d_{\rm DCBH}$) may be the most frequently occurring 
primordial objects in the vicinity of high-z quasars.

\capstartfalse
\begin{deluxetable}{cccccccc}
  \tablecolumns{3}
 \tablewidth{3.0 in}
  \tablecaption{Distances of DCBHs for various effective
    temperatures of the stars in the quasar host galaxy}
  \label{tab_properties}
  \tablehead{     && \multicolumn{2}{c}{ \, \, \, \, \, \, \, \, \,
     Stellar temperature [${\rm K}$]}  \\  \cline{3-6} \\ \colhead{BH mass [M$_{\odot}$]} && \colhead{10$^4$} & \colhead{3 $\times$ 10$^4$} & \colhead{10$^5$}}
\startdata
   10$^8$  &&  300  & 50 &  20 \\      
    10$^9$  && 1000 & 200   &    70    \\
    10$^{10}$  && 3000  & 500  &  200      
\enddata
\tablecomments{Maximum distance $d_{\rm DCBH}$ (in kpc) from host quasars powered
  by black holes with masses $M_{\rm BH}$ in which DCBH can form, for various values of the effective temperature $T_{\rm *}$
  of the stellar population in
  the quasar host halo.  Eddington accretion is assumed for the BH in the source
  quasar in all cases shown here, as is a stellar to black
  hole mass ratio of $M_{\rm *}$/$M_{\rm BH}$ = 100.}
\end{deluxetable}
  \capstarttrue


\capstartfalse
\begin{deluxetable}{cccccccc}
  \tablecolumns{3}
 \tablewidth{3.0 in}
  \tablecaption{Distances of DCBHs for various quasar host galaxy stellar to BH
    mass ratios}
  \label{tab_properties}
  \tablehead{     && \multicolumn{2}{c}{ \, \, \, \, \, \, \, \, 
     Stellar to BH mass ratio}  \\  \cline{3-6} \\ \colhead{BH mass [M$_{\odot}$]} && \colhead{1} & \colhead{10} & \colhead{100}}
\startdata
   10$^8$  &&  10  & 20 & 50  \\      
    10$^9$  && 30 & 70  &  200      \\
    10$^{10}$  && 100  & 200 & 500       
\enddata
\tablecomments{Maximum distance $d_{\rm DCBH}$ (in kpc) from host quasars powered
  by black holes with masses $M_{\rm BH}$ in which DCBH can form, for various ratios of the stellar to black hole mass
  ($M_{\rm *}$/$M_{\rm BH}$) in the quasar host halo.  Eddington accretion is assumed for the BH in the source
  quasar in all cases shown here, as is an effective stellar
  temperature of $T_{\rm *}$ = 3 $\times$ 10$^4$ K.}
\end{deluxetable}
  \capstarttrue


\section{Suppression of Star Formation}

In our picture of bright Pop~III starburst and/or DCBH formation, it
is critical that the massive haloes in which these objects may form
remain metal-free, as even small amounts of heavy elements will
prevent Pop~III star formation and will likely also preclude DCBH
formation (e.g. Omukai et al. 2008).  This implies that star formation
must be suppressed
in them throughout their growth until they reach mass scales of up to $\ga$ 10$^{10}$ M$_{\odot}$, by z $ \simeq$ 6.  Here we
consider the growth history of a typical such halo, which corresponds
to a $\simeq$ 2$\sigma$ overdensity in the cosmological dark matter field
(e.g. Barkana \& Loeb 2001).  We also model the growth of the
photoheated region surrounding a high redshift quasar with a mass that is consistent with that of the highest redshift quasars found to date (Mortlock et al. 2011; Ba{\~ n}ados et al. 2017).  

Figure 7 shows the virial temperature of a 2$\sigma$ halo,
corresponding to a halo with mass 10$^{10}$ M$_{\odot}$ at z = 6, at
three representative redshifts ($z$ = 19, 11 and 7) during its growth.
Also shown are the temperature profiles of the gas in the vicinity of
a growing BH powering the high redshift quasars that are modeled in the
cosmological radiation hydrodynamics simulations of Smidt et
al. (2018), at these same three representative redshifts.      At early times, the gas
is too hot to collapse into the 2$\sigma$ target halo (i.e. T$_{\rm gas}$ $>$ T$_{\rm vir}$; see e.g. Okamoto et
al. 2008), unless it is
farther from the quasar than $\sim$ 100 kpc by z = 7.  Thus, haloes
with masses of $\sim$ 10$^{10}$ M$_{\odot}$ will not form stars in the
vicinity of such high redshift quasars, unless they lie farther than
$\sim$ 100 kpc away by z $\simeq$ 7.  Cosmological dark matter
simulations predict that there are expected to be haloes within this mass
range that lie at distances $d$ $<$ 100 kpc of the $\sim$ 10$^{12}$
M$_{\odot}$ haloes inferred to host the highest redshift quasars (see
e.g. Poole et al. 2017; Poulton et al. 2018).  This implies that star
formation in these haloes may in fact be suppressed as they grow, setting the
stage for a Pop~III starburst or DCBH to form as we have explored
here.  Farther away from the quasar, it may only be lower mass 
(e.g. 10$^8$ - 10$^9$ M$_{\odot}$) haloes in which star formation is suppressed; these objects may still host
DCBHs or Pop~III starbursts (e.g. Trenti et al. 2009; Johnson et
al. 2010; Visbal et al. 2017), although they would likely be less
luminous than those hosted by more massive haloes.

  \begin{figure}
   \includegraphics[angle=0,width=3.5in]{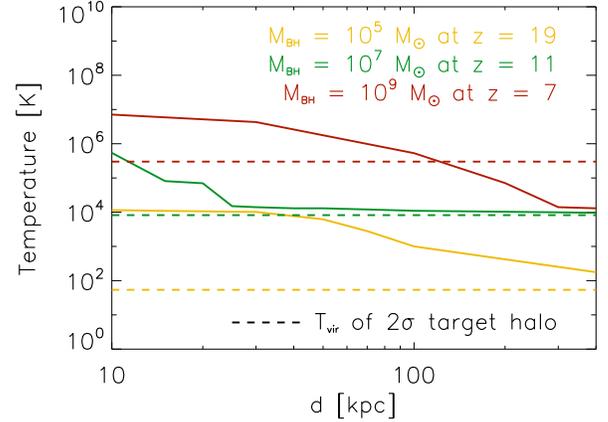}
  \caption{Temperature of the gas ($T_{\rm gas}$) surrounding a canonical high
    redshift quasar powered by a BH that grows to a mass of $M_{\rm
      BH}$ = 10$^9$ M$_{\odot}$ by $z$ = 7 via Eddington-limited
    accretion, as a function of the distance $d$ from the quasar, at three
    epochs of its growth ({\it solid}):  $z$ = 19 ({\it yellow}), 11
    ({\it green})  and 7 ({\it red}); at these times the
    BH mass is 10$^5$, 10$^7$ and 10$^9$ M$_{\odot}$, respectively.
    Also shown is the virial temperature ($T_{\rm vir}$) of a
    2$\sigma$ halo ({\it dashed}), corresponding to a massive
    (10$^{10}$ M$_{\odot}$) potential Pop~III star forming halo at $z$
    $\sim$ 6, at
    the same three epochs.  Star formation would be suppressed in the 
    2$\sigma$ halo up to $z$ = 7 within $\simeq$ 100 kpc, as the gas
    temperature remains higher than its virial temperature.}
   \end{figure}

\section{Detectability and Search Strategy}
One key question is whether or not the primordial clusters or DCBHs
that form in massive halos in the vicinity of high-z quasars are
bright enough to be detected by current or future observational
facilities.  Indeed, this is an especially important question with
regard to the JWST, as it has been designed with the goal of finding
the earliest stellar populations in the high-z universe.  Here we
estimate the luminosities expected for these primordial objects and
address the prospects for uncovering them with the JWST.

The brightest emission line expected from high-z primordial star
clusters is the hydrogen Lyman $\alpha$ line, produced in the H$~{\rm II}$
regions around hot massive stars (e.g. Schaerer 2002).  As shown
in Figure 8, the NIRCam survey instrument on board the JWST provides an exquisite
sensitivity to this emission line at the high redshifts where we
expect primordial stellar clusters and DCBHs to form.\footnote{https://jwst.stsci.edu/instrumentation/nircam}  
Along with flux limits for NIRCam, Figure 8 shows estimates of 
the Lyman $\alpha$ fluxes expected from Pop III stellar clusters, as
functions of their host halo mass expressed in terms of virial
temperature and the assumed stellar initial mass function (IMF).
These fluxes are computed following the approach adopted in Johnson
(2010), but using an updated estimate for the star formation efficiency
of $f_{\rm *}$ = 0.002 that is consistent with the values gleaned from cosmological radiation
hydrodynamics simulations of Pop III star formation (e.g. Aykutalp et
al. 2014; Xu et
al. 2016; Barrow et al. 2018).  Although this conservative choice for
the star formation efficiency strongly limits the Lyman $\alpha$ flux,
the predicted fluxes are still high enough to be detected with at
least 3
sigma confidence from clusters formed in sufficiently massive halos, in particular from
halos in the upper end of the mass range we predict to host Pop III
star formation in the vicinity of high-z quasars, with a 2 $\times$
10$^5$ sec exposure as shown in Figure 8.   

\begin{figure}
  \includegraphics[angle=0,width=3.4in]{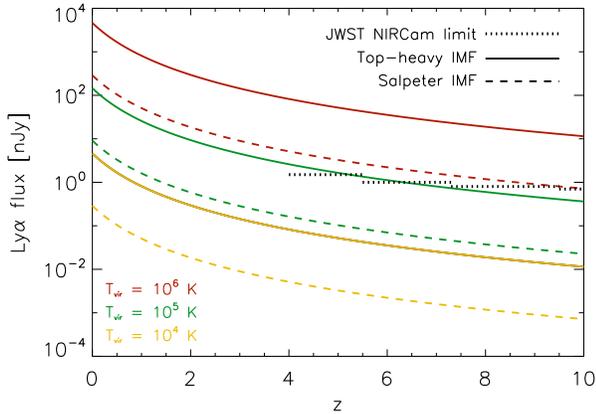}
  \caption{The Lyman $\alpha$ fluxes predicted for Pop III star
    clusters formed in halos with virial temperatures of T$_{\rm vir}$
    = 10$^4$ ({\it
      yellow lines}), 10$^5$ ({\it
      green lines}), and 10$^6$ K ({\it
      red lines}), assuming a relatively low star formation efficiency
    of $f_{\rm *}$ = 0.002, for a Salpeter IMF ({\it dashed lines})
    and a top-heavy IMF characterized by stars with masses of order 100
    M$_{\odot}$ ({\it solid lines}).  The flux limits for the NIRCam survey
    instrument aboard the JWST for a 3 sigma detection and exposure
    time of 2 $\times$ 10$^5$ sec are also shown by the dotted lines.
    Pop III clusters formed in the massive halos (T$_{\rm vir}$ $\ga$
    10$^5$ K) expected to
    potentially host primordial star formation in the vicinity of
    high-z quasars are expected to be bright enough for detection by JWST.
}
\end{figure}

\begin{figure*}
  \includegraphics[angle=90,width=7.1in]{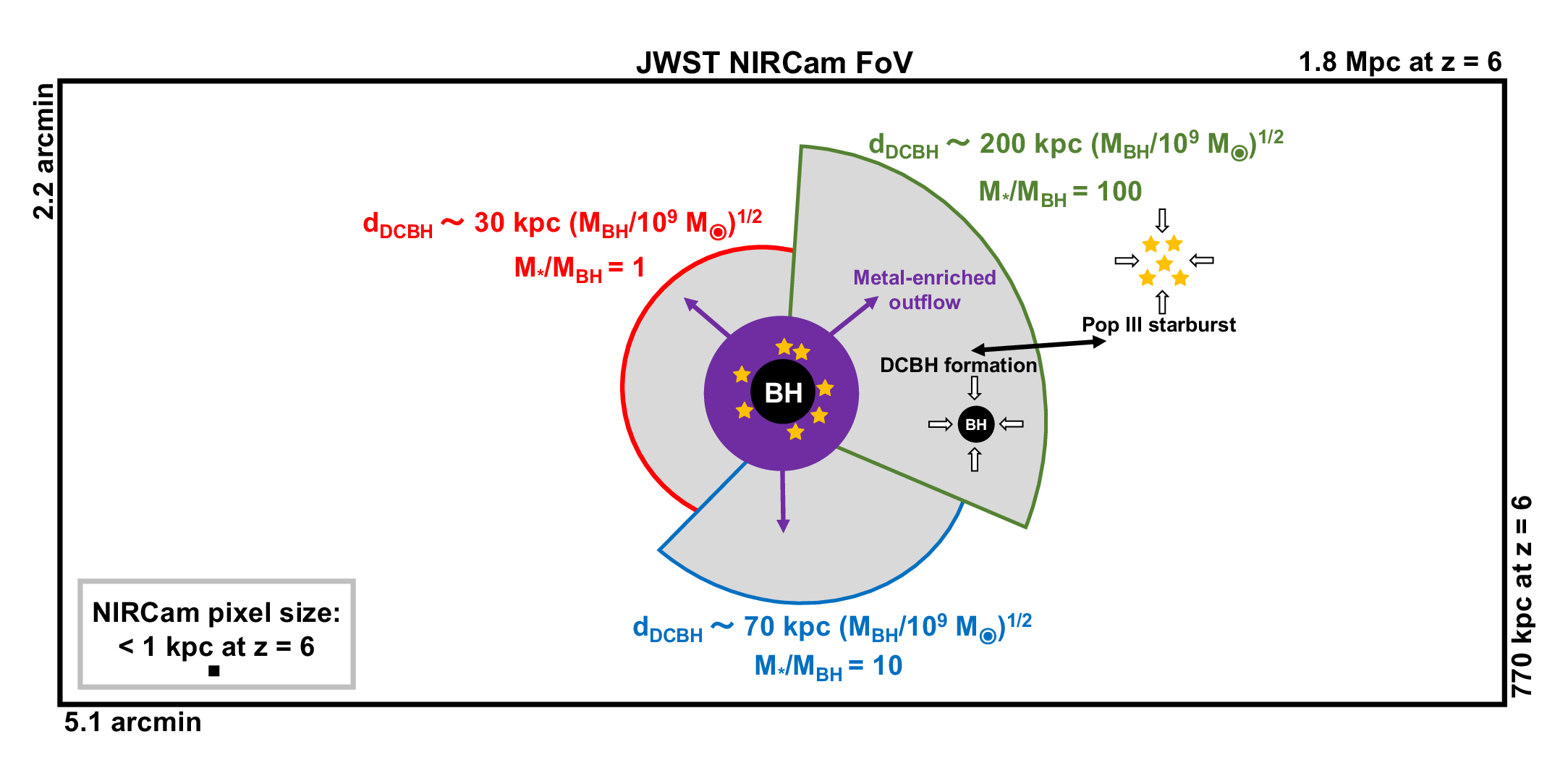}
  \caption{Graphical representation of our findings presented in Table
    2 for the maximum distance $d_{\rm DCBH}$ from a high redshift quasar at which a
    DCBH can form in a collapsing primordial halo, as it depends on
    the mass $M_{\rm BH}$ of the black hole powering the quasar and on
    the stellar mass $M_{\rm *}$ in the quasar host galaxy.  
In general, DCBH formation can occur farther away from quasars powered
by more rapidly accreting black holes, as it can also from quasar host
galaxies containing a larger mass in stars.  In general, the region
surrounding a $z$ $\ga$ 6 quasar within which bright Pop III clusters or
DCBHs are expected to form lies well within the field of view (FoV) of
the NIRCam survey instrument aboard the JWST, shown here for
reference.  Thus, potentially promising surveys for bright Pop III
clusters or DCBHs could be carried out
using single pointings of NIRCam at known high-z quasars.
}
\end{figure*}

One of the fortuitous aspects of the formation sites of the
potentially brightest Pop III clusters is that, being near bright quasars, they
can be relatively easily identified.  As we have shown, the
brightest Pop III clusters and DCBHs are expected to form
within several hundred kiloparsecs from high-z quasars.  As shown schematically in
Figure 9, these regions lie well within the field of view of the JWST
NIRCam survey instrument, which covers roughly a square megaparsec at $z$
$\ga$ 6.  This suggests that a straightforward search strategy
by which JWST may have a chance to detect clusters of Pop III stars is
to use single pointings of NIRCam at individual known high-z quasars.
Follow-up of detected sources may then be done using the NIRSpec instrument to
obtain spectra which could identify candidate Pop III
clusters or DCBHs based on bright helium emission lines and
constraints on lines from heavy (non-primordial) species such as
carbon and oxygen (e.g. Natarajan et al. 2017; Barrow et al. 2018).  Given the distance of these sources from nearby
quasars, the accretion luminosity from the quasar and the stellar mass
in its host galaxy, equation (3) can also be used as an additional 
baseline to distinguish between Pop III clusters and DCBHs.\footnote{The number density of DCBH seeds in general and of DCBH
  seeds formed in halos irradiated by quasars at high redshift is not
  well-constrained, in part due to its sensitive dependence on the
  radiative flux required for their formation (e.g. Habouzit et
  al. 2016).  Recent work, however, shows that they may form readily
  due to dynamical heating in rapidly growing atomic cooling halos
  (e.g. Wise et al. 2019).}  
Short of relying on gravitational lensing to
magnify the light from less massive Pop III clusters (e.g. Rydberg et
al. 2015; Windhorst et al. 2019) or identifying Pop III supernovae
(e.g. Weinmann \& Lilly 2005; Frost et al. 2009; de Souza et al. 2014; Hartwig et al. 2018), this is perhaps the most promising
strategy for finding Pop III stars with the JWST.

\section{Summary}
We have explored the evolution of the primordial gas as it is exposed to
the extreme radiation emitted from high-z quasars powered by 
rapidly accreting supermassive black holes and by their host stellar population.  As shown in Figure 2, we
confirm that the temperature of the gas is raised to values of
up to $\sim$ 10$^6$ K due to the intense X-ray flux, with the implication that the gas will only collapse once it
has been incorporated in DM haloes with masses up to $\sim$ 10$^{11}$
M$_{\odot}$.  Such large haloes would provide extremely large mass
reservoirs of gas from which Pop~III stars could form, setting the
stage for the brightest primordial starbursts in the early Universe.  

The possible final outcomes of the collapse of the gas that we find are shown
schematically in Figure 9.   The intense LW radiation that is emitted from quasar
host galaxies suppresses the cooling of the primordial gas as
it collapses, thus satisfying the requirements for the formation of
DCBHs in close proximity to high-z quasars.  Farther
out, bright Pop~III starbursts are instead likely to occur, with the
mass function of stars likely shifting to lower masses farther from
the quasar, in part due to the action of HD cooling and in part due to
the more intense heating of the gas by X-rays closer to the quasar
(see also e.g. Hocuk \& Spaans 2010).

It is key that predictions for the sites of primordial star and seed black
hole formation be put forward now, as they are set to be tested in the
upcoming years with the launch of the JWST, among other next
generation facilities that will peer deeper than ever into the distant
early Univserse.  This work provides a clear prediction that the
brightest Pop~III starbursts could be located within $\sim$ 1 Mpc of bright
quasars at $z$ $\ga$ 6, and that DCBHs may be formed in even closer
proximity to them.\footnote{ We note that it may be only a small
    fraction of all DCBHs that form in the vicinity of high redshift
    quasars (see e.g. Agarwal et al. 2012; Yue et al. 2014; Habouzit et
    al. 2016).}  While these primordial objects are likely rare,
it follows from this prediction where they should be sought:  near 
bright quasars in the early Universe, perhaps even around some 
that have already been discovered.  Indeed, we have shown that the
brightest Pop III clusters may be detectable by the  JWST, and that a
potentially promising strategy for finding them would be to make
single pointings at individual known high-z quasars with the NIRCam
survey instrument.

In addition to future observations
that will place constraints on the number density of such objects in
the early Universe, next generation cosmological radiation
hydrodynamics simulations will also be able to place theoretical
constraints (see e.g. Habouzit et al. 2018).  Such simulations will be challenging, as they will 
have to cover a large cosmological volume of at least hundreds of
comoving Mpc in order to contain even one quasar like the earliest known
at $z$ $\ga$ 7 and at the same time have sufficiently high resolution
to model star formation in minihaloes and the impact of X-rays and LW
radiation from the quasar on the gas within at least several comoving
Mpc of its host galaxy (currently such simulations are limited to much smaller
computational volumes; see e.g. Maio et al. 2018).  Whether these next generation simulations can
be completed before the beginning of the next generation of
observations with JWST remains to be seen.




\section*{Acknowledgements}
Work at LANL was done under the auspices of the National Nuclear Security 
Administration of the US Department of Energy at Los Alamos National 
Laboratory under Contract No. DE-AC52-06NA25396.  J. ~L. ~J.
and A.~A. are  supported  by  a  LANL  LDRD
Exploratory  Research  Grant  20170317ER.  This work was initiated and
performed in part at Aspen Center for Physics, which is supported by
National Science Foundation grant PHY-1607611.  We are grateful to an
anonymous reviewer for comments which led to improvements in the
presentation of this work, as well as to Zoltan Haiman, Mihir
Kulkarni, Eli Visbal, Kohei Inayoshi and Kevin Hainline for valuable discussions.



\end{document}